# Influence of magnetic entropy on stability and thermodynamic properties of Laves phase Fe$_2$Mo from first principles

**Dmitry Vasilyev**

Baikov Institute of Metallurgy and Materials Science of RAS, 119334, Moscow, Leninsky Prospekt 49, Russia

*dvasilyev@imet.ac.ru; vasilyev-d@yandex.ru*

**Abstract**. The thermodynamic and physical properties of the Laves phase Fe$_2$Mo have been investigated using the finite-temperature quantum mechanical calculations within the frame of the density functional theory (DFT). All relevant free energy contributions including electronic, vibrational and magnetic excitations are considered. The quasi-harmonic Debye - Grüneisen theory is used. In this work, a new method of searching a thermal expansion path of compounds is proposed. It allows to reduce the problem to a one-dimensional case and minimize the free energy in one variable, in volume. The heat capacity, thermal expansion, elastic constants and bulk modulus are modelled. The calculated results analyzed and are in an agreement with the available experimental data. It is shown that magnetic entropy must be considered on equal footing with vibrational and electronic energies to reliably predict stability of Fe$_2$Mo.

## 1. Introduction

The Laves phase Fe$_2$Mo is one of the intermetallic phases, the existence of which is taken into account when developing ferritic steels. Ferritic steels have good corrosion resistance, low creep, and low swelling under irradiation and at elevated temperatures [1, 2], therefore, they present both theoretical and practical interest, and considered as promising for use as fuel element cladding of the new generation reactors and as the first wall of future fusion reactors. The Laves phase Fe$_2$Mo (λ-) and other the μ-, R-, σ- and γ- phases exist in the Fe-Mo phase diagram, which is one of the main systems for the development of ferritic steels. These phases essentially limit the stabile area of bcc solid solutions. The exact positions of these phase boundaries on the Fe-Mo phase diagram is a necessary condition for the successful design of materials used in industry. In construction steels, the Laves phase is a hardening phase, which precipitated during heating or operation of a product for a long time at high temperature. A characteristic feature of the Laves phase is their slow precipitation from solid solution and the slow growth of particle sizes, which favourably affects the stabilization of the mechanical properties that occur during long isothermal exposures [3]. The precipitation of Laves phase in ferritic steels can improve tensile, creep and oxidation properties.

In order to understand the kinetic processes taking place in ferritic steels the knowledge of binary and ternary (for example, Fe-Cr-Mo) phase diagrams as well as the thermodynamic properties for the Fe$_2$Mo phase is essential. Obtaining thermodynamic properties at finite temperatures from experiment is a difficult task. This is due to the fact that the Laves phase Fe$_2$Mo is located in the low-temperature part of the phase diagram, where diffusion processes proceed slowly. And the presence of other phases, such as μ-, R- and σ-, which force the experimenters to make additional efforts to prevent their precipitations. To understand the diffusion and kinetic processes occurring in ferritic steels, it is advisable to carry out computer simulation to obtain Gibbs potentials.

Quantum-mechanical calculations can help in calculating the total energy of a compound in the ground state (at T = 0 K) and in obtaining the temperature dependence of the Gibbs potential. To obtain them, it is necessary to calculate the equilibrium parameters of the crystal lattice, the elastic constants $C_{ij}$ of the strain tensor, and the Debye temperature of the compound in the ground state. Then, apply some model to calculate the thermodynamic properties at finite temperatures.

Several studies on the structural properties and lattice stabilities of $Fe_2Mo$ have been conducted over the past years. The lattice parameters, the formation enthalpy and thermodynamic functions of Fe-Mo alloys were reported in the work [5] using CALPHAD approach and Density Functional Theory (DFT). Lattice parameters, optimized atoms positions and total magnetic moments of $Fe_2Mo$ were obtained in the work [6] with the help of the DFT calculations. Bulk modulus and lattice parameters and of $Fe_2Mo$ were predicted by DFT in the work [7]. From above, one could see that these studies are mainly focusing on structure properties of $Fe_2Mo$. However, despite a great importance of developing the ferrite steels, the elastic, thermodynamic and physical properties of $Fe_2Mo$ are not well compared and reported at the same time. Moreover, as far as is known, the magnetic subsystem and their influence on the stability of Laves phase $Fe_2Mo$ have not yet been investigated. Nevertheless, magnetic entropy can contribute up to one third to free energy of a compound, as was shown in the work [9]. Since these data are essential for calculating Gibbs energy and obtaining the thermodynamic and physical properties of alloys at finite- temperatures, their absence was a motivation prompted to study them.

In this work, the structure, elastic, thermodynamic and physical properties of Laves phase $Fe_2Mo$ in the ground state (at T = 0 K) and at finite-temperatures were studded by quantum mechanical calculations. The structure is optimized by full relaxation procedure and the lattice parameters are obtained. The heat of formation energy is calculated. The elastic constants $C_{ij}$ of the strain tensor were obtained using the distortion matrices that deform the lattice and by calculating the corresponding changes in the total energies. The Debye temperature and sound velocities in different crystallographic directions were predicted. A new method allowing to avoid differentiation in many variables and obtain a path of thermal expansion and contraction by comparing the free energies calculated along different paths, is developed.

The quasi-harmonic Debye - Grüneisen theory is used to calculate the vibrational contribution to the free energy $F(T)$ of $Fe_2Mo$. The distributions of magnetic moments of atoms on sub-lattices of $Fe_2Mo$ and electronic energy are taken into account as well, in order to investigate the thermodynamic properties as temperature functions. The influence of the magnetic and vibrational subsystems on the stability of $Fe_2Mo$ are analysed. Finally, the thermal expansion $V(T)$, heat capacity $C_p(T)$, elastic constants $C_{ij}(T)$, bulk $B(T)$, shear $G(T)$ and Young's $E(T)$ modulus are predicted at finite temperatures in this work. The obtained results may provide a valuable estimation for the properties which very difficult to obtain in experiments.

## 2. Methodology

The Gibbs potential of a compound, which is necessary for studying diffusion and kinetic processes occurring in steels, as well as for calculating the phase diagram and evaluating thermodynamic properties, can be obtained by transforming the Helmholtz free energy $F(V,T)$ by the Legendre transformation, calculated in the adiabatic form according to [8, 9, 10]:

$$F(V,T) = E_{tot}(V) + F_{el}(V,T) + F_{vib}(V,T) + F_{mag}(V,T) - TS_{conf}(T) \quad (1)$$

where $E_{tot}(V)$ is the ground state total energy calculated by quantum mechanical calculations at T = 0 K which may by fitted using the Birch-Murnaghan equation of state [11]. $F_{el}(V,T)$, $F_{vib}(V,T)$, $F_{mag}(V,T)$ are the electronic, vibrational and magnetic free energies contributions. $S_{conf}(T)$ is the ideal configurational entropy. Description of details of those energy contributions to the Helmholtz free energy $F(V,T)$ are in the below sections.

*2.1. Electronic free energy*

The electronic energy contribution $F_{el}(V,T)$ to the Helmholtz free energy computed based on the quantum mechanical calculations and formulated as in [8, 14]

$$F_{el}(V,T) = E_{el}(V,T) - TS_{el}(V,T), \qquad (2)$$

with $E_{el}$ is given by [12]

$$E_{el}(T,V) = N_A \int_{-\infty}^{\infty} n(\varepsilon,V) f(\varepsilon,T)\varepsilon d\varepsilon - N_A \int_{-\infty}^{\varepsilon_F} n(\varepsilon,V)\varepsilon d\varepsilon \qquad (3)$$

where $n(\varepsilon,V)$ is the total DOS, $f(\varepsilon,T)$ is the Fermi-Dirac distribution and $N_A$ is Avogadro constant. The electronic entropy $S_{el}$ is calculated as follow [13]

$$S_{el}(T,V) = N_A \int_{-\infty}^{\infty} n(\varepsilon,V)\big(f(\varepsilon,T)\ln f(\varepsilon,T) + (1 - f(\varepsilon,T))\ln(1 - f(\varepsilon,T))\big)d\varepsilon \qquad (4)$$

*2.2. Vibrational free energy*

The vibration free energy, $F_{vib}(V,T)$, is described by using the Debye- Grüneisen model described in the work [8], which has been recently applied to investigate the thermodynamics and phase formation in high entropy alloys contained species with magnetic properties [9, 10].

$$F_{vib}(V,T) = E_D(V,T) - TS_{vib}(V,T), \qquad (5)$$

with $E_D$ and $S_{vib}$ given by

$$E_D(T,V) - E_0 = 3N_A k_B T D\left(\frac{\theta_D}{T}\right) \qquad (6)$$

$$S_{vib}(T,V) = 3N_A k_B \left[\frac{4}{3} D\left(\frac{\theta_D}{T}\right) - \ln(1 - \exp\left(-\frac{\theta_D}{T}\right))\right] \qquad (7)$$

where $D(\theta_D/T)$ is the Debye function, $\theta_D$ is the Debye temperature.

$$E_0 = \frac{9}{8} N_A k_B \theta_D \qquad (8)$$

Where $E_0$ is the zero-point energy expressed within the Debye approximation and $k_B$ is the Boltzmann constant.

The volume dependence of the Debye temperature is evaluated through the Grüneisen parameter $\gamma$ [8]

The Grüneisen constant $\gamma$ is calculated as follow [15]:

$$\gamma = -1 - \frac{V}{2}\frac{\partial^2 P/\partial V^2}{\partial P/\partial V} \qquad (9)$$

The Debye temperature $\theta_D$ as a function of volume is calculated as follow [8]:

$$\theta_D = \theta_{Do}\left(\frac{V_0}{V}\right)^\gamma \qquad (10)$$

Where $\theta_{D0}$ is the Debye temperature calculated at T = 0 K, $V_0$ is the lattice equilibrium volume. The Debye temperature $\theta_D$ is derived from the elastic constants and will be described in detail below.

## 2.3. Magnetic free energy

In this work the assumption is made that the contribution to the free energy of the Laves phase $Fe_2Mo$ due to magnetic ordering $F_{mag}(V,T)$ may be described by using the following formula:

$$F_{mag}(V,T) = [F'_{mag}(T) - F'_{mag}(0K)] - TS_{mag}(V) \quad (11)$$

Where $F'_{mag}(T)$ is the magnetic contribution to the free energy according to Hillert and Jarl [16, 17]:

$$F'_{mag}(T) = RT Ln(\beta + 1)f(\tau) \quad (12)$$

where $\tau$ is $T/T_c$, $\beta$ is the average magnetic moment for the $Fe_2Mo$ in Bohr magnetons, $\beta = 0.371$ (per atom, $\mu_B$), as following from the present calculations, see Table 3. $T_c$ is the Curie temperature, the function $f(\tau)$ is defined below and above $T_c$ according the works [16, 17]. The difference $(F'_{mag}(T) - F'_{mag}(0K))$ in (11) is due to the constant contribution of magnetic internal energy which is included into the total ground energy.

The theoretical value of Curie temperature $T_c$ was estimated by employing the mean-field approximation according the work [9]

$$T_C = \frac{2}{3k_B}(E_{tot}^{PM}(V) - E_{tot}^{FM}(V)) \quad (13)$$

Where $E^{PM}_{tot}(V)$ is the ground state total energy of the paramagnetic (PM) state and $E^{FM}_{tot}(V)$ is the ground state total energy of the ferromagnetic (FM) state of a compound, these energies were calculated at the corresponding equilibrium atomic volumes.

The magnetic entropy, $S_{mag}(V)$, and configurational entropy, $S_{conf}(T)$, are treated by mean-field approximation:

$$S_{mag}(V) = N_A k_B \sum_{i=1}^{n} c_i ln(|\mu_i(V)| + 1) \quad (14)$$

$$S_{conf}(T) = N_A k_B \sum_{i=1}^{n} c_i ln\, c_i \quad (15)$$

where $\mu_i$ is the local magnetic moment of atom $i$, $c_i$ is the atomic concentration. $N_A$ is Avogadro and $k_B$ is Boltzmann constants.

## 2.4. Elastic properties

The elastic constants $C_{ij}$ of the strain tensor determine the response of the crystal lattice distorted by applied forces, and characterize the bulk modulus $B$, shear modulus $G$, Young's modulus $E$, and Poisson's ratio $v$. So $C_{ij}$ play an important part in the mechanical properties of material science. For example, the Debye temperature was calculated in this work through the elastic constants.

In the calculations of the elastic constants $C_{ij}$, the technique of distortion matrices $D_i$ was used. When distortion matrices are applied to a lattice, its shape is slightly distorted, which means that the lattice parameters and angles lying between the unit vectors of the crystal lattice change. Those small distortions simulate the necessary external small stress which needed to be applied to the equilibrium lattice in order to determine the corresponding variation in the total energy. Then, through this variation in the total energy, the elastic constants $C_{ij}$ can be calculated using the elastic strain energy which was given as follows [18]

$$U = \frac{\Delta E}{V_0} = \frac{1}{2}\sum_{i}^{6}\sum_{j}^{6} C_{ij} e_i e_j \quad (16)$$

where $\Delta E = E_{tot}(V, \delta) - E_{tot}(V_0, 0)$ is the total energy difference between deformed crystal lattice and the initial lattice, $V_0$ is the volume of lattice at equilibrium state and $C_{ij}$ are the elastic constants; $e_i$ and $e_j$ are strains. And $\delta$ is the deformation added to the equilibrium lattice, it defines the output value of the distortion imposed on the crystal lattice by applying a distortion matrix. Then, the elastic constants $C_{ij}$ can be identified from graphs like those shown in Figure 2 as proportional to the second order coefficient in a polynomial fit of the total energy as a function of the distortion parameter $\delta$ [18].

In order to use the technique of distortion matrices $D_i$, the Laves phase Fe$_2$Mo lattice was described in a matrix form by Bravais lattice vectors of a hexagonal crystal lattice which has two parameters $a$ and $c$, and in a matrix form those vectors are defined as

$$R = \begin{pmatrix} \frac{\sqrt{3}}{2} & \frac{1}{2} & 0 \\ -\frac{\sqrt{3}}{2} & \frac{1}{2} & 0 \\ 0 & 0 & 1 \end{pmatrix} \quad (17)$$

The elastic constants $C_{ij}$ were determined by applying the symmetric distortion matrices $D_i$, which contain the strain components $\delta$, to the matrix with initial lattice vectors $R$ (17) according to the relation $R \cdot D_i = R'$, where $R'$ is the deformed matrix containing distorted lattice vectors. So, by calculating $\Delta E$ of distorted lattice one can calculate the elastic constants. Due to the fact that, that a hexagonal structure has five independent elastic constants ($C_{11}$, $C_{12}$, $C_{13}$, $C_{33}$, $C_{44}$) and one dependent constant which is calculated as $C_{66} = (C_{11}-C_{12})/2$, according with the work [19], the five different strains are needed to determine them. The five distortion matrices $D_i$ used to simulate the strains on the Fe$_2$Mo lattice, in the present calculations, are given in Table 1 with the corresponding formulas of changes in the energy as a function of applied strain. The feature of the $D_2$, $D_3$ and $D_5$ matrices is when they are used to distort the lattice, the volume remains constant, while, when the other two matrices are distorting the lattice its volume changes. The technique utilizing these distortion matrices approach for calculating the elastic constants $C_{ij}$ of the strain tensor have been described in the work [20].

**Table 1.** Distortion matrices for deforming the lattice of Fe$_2$Mo and corresponding changes in total energies $\Delta E(\delta)$ used for calculations of elastic constants $C_{ij}$.

| Distortion matrix | Energy change due to applied strain |
|---|---|
| $D_1 = \begin{pmatrix} 1+\delta & 0 & 0 \\ 0 & 1 & 0 \\ 0 & 0 & 1 \end{pmatrix}$ | $\Delta E = V_0(\tau_1 \delta + \frac{C_{11}}{2}\delta^2)$ |
| $D_2 = \begin{pmatrix} \frac{1+\delta}{(1-\delta^2)^{1/3}} & 0 & 0 \\ 0 & \frac{1-\delta}{(1-\delta^2)^{1/3}} & 0 \\ 0 & 0 & \frac{1}{(1-\delta^2)^{1/3}} \end{pmatrix}$ | $\Delta E = V_0\left[(\tau_1 - \tau_2)\delta + \frac{1}{2}(C_{11} + C_{22} - 2C_{12})\delta^2\right]$ |
| $D_3 = \begin{pmatrix} \frac{1+\delta}{(1-\delta^2)^{1/3}} & 0 & 0 \\ 0 & \frac{1}{(1-\delta^2)^{1/3}} & 0 \\ 0 & 0 & \frac{1-\delta}{(1-\delta^2)^{1/3}} \end{pmatrix}$ | $\Delta E = V_0\left[(\tau_1 - \tau_3)\delta + \frac{1}{2}(C_{11} + C_{33} - 2C_{13})\delta^2\right]$ |
| $D_4 = \begin{pmatrix} 1 & 0 & 0 \\ 0 & 1 & 0 \\ 0 & 0 & 1+\delta \end{pmatrix}$ | $\Delta E = V_0(\tau_3 \delta + \frac{C_{33}}{2}\delta^2)$ |
| $D_5 = \begin{pmatrix} \frac{1}{(1-\delta^2)^{1/3}} & 0 & 0 \\ 0 & \frac{1}{(1-\delta^2)^{1/3}} & \frac{\delta}{(1-\delta^2)^{1/3}} \\ 0 & \frac{\delta}{(1-\delta^2)^{1/3}} & \frac{1}{(1-\delta^2)^{1/3}} \end{pmatrix}$ | $\Delta E = V_0(2\tau_4 \delta + 2C_{44}\delta^2)$ |

*2.5. DFT calculation*

The calculations were performed by the first principles calculations implemented in the WIEN2k package [21], which employs the Full Potential-Linear Augmented Plane Wave (FP-LAPW) Method [21] based on the density functional theory (DFT) [22, 23]. The exchange and correlation terms were described by the generalized gradient approximation (GGA) with the Perdew-Bruke-Eruzerhof (PBE) functional [24]. The PBE functional is reliable in describing magnetic materials and was used in the works [9, 10, 14]. The parameters $RK_{max}$ (the minimum radius times the maximum plane-wave coefficient) and muffin-tin radius ($R_{MT}$) were set as $R_{MT}$ = 2.05 for Fe, $R_{MT}$ = 2.15 for Mo and $RK_{max}$ = 9. To model the $Fe_2Mo$ compound the cell of 12 atoms was used with a 13 x 13 x 6 $k$-point mesh in the first irreducible Brillouin zone using Monkhorst-Pack scheme [25]. To confirm the convergence of the calculations, a careful investigation of the dependence of the total energy on the size of the grid for the $k$-point mesh was carried out. The structure optimizations were performed by full relaxation until the maximum force on the atom was below 5 meV/Å and the maximum stress was below 0.02 GPa. The calculations of the total energy and electronic structure were accompanied by cell optimization with permissible variation of a self-consistent field (SCF) less than $1 \times 10^{-6}$ eV/atom.

## 3. Results and discussion.

### 3.1. Ground state properties

#### 3.1.1. Crystal structure and lattice constants

The Laves phase $Fe_2Mo$ is a stoichiometric compound with the fixed composition and possess the hexagonal $MgZn_2$ structure, with space group *P6_3/mmc* (No. 194) and Pearson symbol is hP12, as shown in Figure 1. It contains 8 Fe atoms and 4 Mo atoms in the unit cell, where the Fe atoms occupy the 2*a* Wyckoff site (0, 0, 0) and 6*h* site (0.830, 0.660, 0.250), Mo atoms occupy the 4*f* site (0.333, 0.667, 0.063).

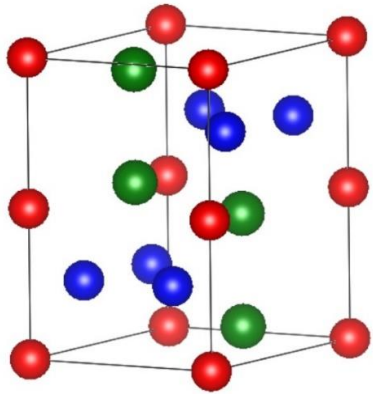

**Figure 1.** Unit cell of C14- Laves phase $Fe_2Mo$, Fe atoms occupy 2*a* and 6*h* Wyckoff sites are shown in red and blue respectively, Mo atoms occupy 4*f* sites are shown in green. The lattice structure drawing was prepared by the Vesta-3 software [4].

The ground-state properties of $Fe_2Mo$ were obtained by the spin-polarized calculations with the geometry optimizations and the full structural relaxation optimization, which have been performed from the above configuration. The optimized lattice parameters are listed in Table 2 and in Figure 3, together with the available experimental data [7, 26, 27] and other theoretical results [5, 6, 7, 28, 29, 30] for comparison. It should be noted that the obtained parameters of the equilibrium lattice are in a satisfactory agreement with the experimental data and lie within 1.5%. The calculated enthalpy formation of the Fe$_2$Mo phase is negative meanings that the compound is stable, the obtained value is shown in Table 2 together with the other theoretical values. The obtained optimized atomic coordinates are given in Table 3. Their comparison with the theoretical results [6] is showing a satisfactory agreement within 1 %. The local magnetic moments are reported in Table 3 as well, the calculated total magnetic moment in the cell (including the interstitial space) is 4.451 ($\mu_B$).

**Table 2**
Ground state properties of Fe$_2$Mo at T=0K and available experimental data.

|              | $a$ (Å) | $c$ (Å) | $c/a$ | $V$ (Å$^3$) | $\Delta H$ (kJ/mol) |
|---|---|---|---|---|---|
| This work    | 4.672 | 7.775 | 1.664 | 147.004 | -2.210 |
| Exp. [7]     | 4.745 | 7.734 | 1.630 | 150.802 | |
| Exp. [27]    | 4.73  | 7.72  | 1.632 | 149.579 | |
| Exp. [26]*   | 4.731 | 7.768 | 1.642 | 150.573 | |
| This work**  | 4.760 | 7.714 | 1.621 | 151.353 | |
| This work*** | 4.728 | 7.868 | 1.664 | 152.309 | |
| Calc. [6]    | 4.659 | 7.743 | 1.662 | 145.585 | |
| Calc. [7]    | 4.682 | 7.560 | 1.615 | 143.551 | |
| Asses. [29,30] | 4.744 | 7.725 | 1.628 | 150.563 | |
| Calc. [28]   |       |       |       |         | -1.534 |
| Calc. [5]    | 4.669 | 7.793 | 1.669 | 147.124 | -0.714 |

\* obtained at T = 1073 K after 1500 hours of annealing.
\*\* calculated in this work at T = 1073 K by applying the quasi-harmonic Debye - Grüneisen theory with allowance for the distributions of the magnetic moments of atoms on the sublattices.
\*\*\* calculated in this work at T = 1073 K by applying the quasi-harmonic Debye - Grüneisen theory without accounting magnetic contributions to the free energy.

**Table 3**
Obtained ground state data of atomic coordinates of Fe$_2$Mo and local magnetic moments per atom.

| species | site | x | y | z | moment ($\mu_B$) |
|---|---|---|---|---|---|
| Fe | 2a | 0.00000 | 0.00000 | 0.00000 | -1.241 |
| Fe | 6h | 0.17050 | 0.34100 | 0.75000 | +1.195 |
| Mo | 4f | 0.33333 | 0.66667 | 0.06978 | -0.034 |

*3.1.2. Calculation of elastic constants*

The lattice of Fe$_2$Mo was deformed by successively imposing on it distortion matrices $D_i$ listed in Table 1. The corresponding energy dependences are obtained, which are necessary for calculating the required elastic constants. For each type of lattice distortions, the total energy was calculated for various deformations $\delta = \pm\, 0.01i$. The variations in total energies $\Delta E(\delta)$ as functions of applied strains for the different types of distortions $D_i$ of Fe$_2$Mo lattice are plotted in Figure 2. The values of the second order coefficients which match the ($C_{11}/2$, ($C_{11}+C_{22}-2C_{12}$)/2, ($C_{11}+C_{33}-2C_{13}$)/2, $C_{33}/2$ and $2C_{44}$) parts of the sum of corresponding equations listed in Table 1, were accordingly calculated. Then, the five elastic constants $C_{ij}$ were calculated by solving the corresponding set of equations. The calculated results of the ground state elastic constants $C_{ij}$ for Fe$_2$Mo compound are listed in Table 4. There aren't available experimental or theoretical data of the elastic constants for Fe$_2$Mo in literature, therefore, the obtained result may be of interest to experimenters.

The generalized criterion for the mechanical stability of hexagonal crystals at zero pressure imposes the following restrictions on the elastic constants, according to the work [19]:

$$C_{11} > 0;\ (C_{11} \cdot C_{33} - 2 \cdot C_{13}^2 + C_{12} \cdot C_{33}) > 0;\ C_{11} - |C_{12}| > 0;\ C_{44} > 0 \qquad (18)$$

**Table 4**
Ground state elastic constants $C_{ij}$ (GPa) of single crystal Fe$_2$Mo compound at T=0K.

| $C_{11}$ | $C_{12}$ | $C_{13}$ | $C_{33}$ | $C_{44}$ |
|---|---|---|---|---|
| 459.27 | 170.38 | 105.49 | 379.00 | 113.73 |

The calculated values of the elastic constants given in Table 4 show that this criterion is met, thus the Laves phase Fe$_2$Mo is a mechanically stable intermetallic at zero pressure. In a crystal lattice the change in stiffness in relation to basic deformations is described by values of $C_{11}$ and $C_{33}$ elastic constants. The shear resistance in the {100} plane and in the <110> direction is reflected by values of $C_{66}$ elastic constant. The value of $C_{44}$ characterises a shear resistance in the {010} or {100} plane in the <001> direction of a lattice. The $C_{11}$ = 459.3 (GPa) and $C_{33}$ = 379 (GPa) values of Fe$_2$Mo are relatively large and they show that it will require a certain amount of force to compress the lattice along *a*- or *c*- axis.

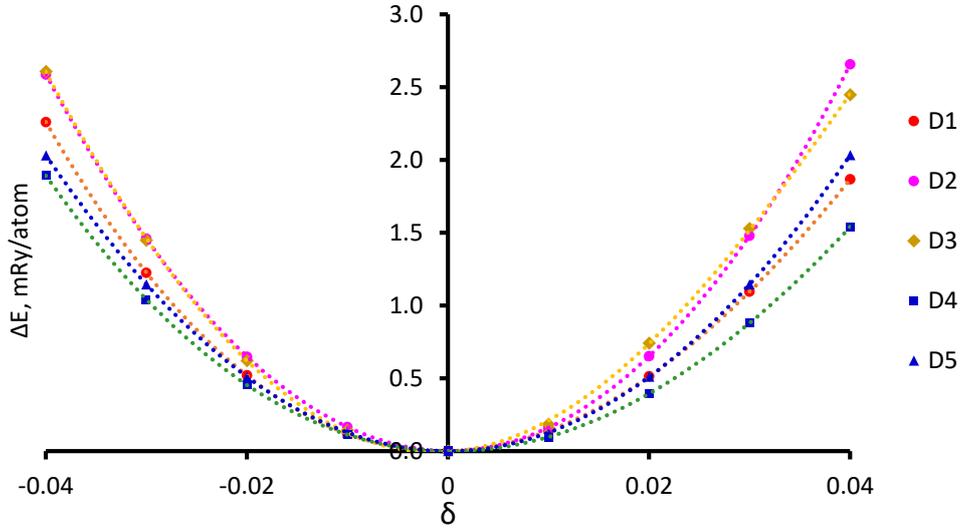

**Figure 2**. Variations in the total energy ($\Delta E= E_{tot}(V, \delta_i)-E_{tot}(V_0,0)$, mRy/atom) vs. strains ($\delta$) induced by imposing the distortion matrices $\boldsymbol{D_1 \div D_5}$ on the Laves phase Fe$_2$Mo lattice. The dotted lines are the polynomial fits.

But, the mechanical properties of materials are usually described by elastic properties of a polycrystalline and not a monocrystalline, for a practical reason. Those polycrystalline elastic properties are bulk modulus *B*, shear modulus *G*, Young's modulus *E* and Poisson's ratio *v*. They are estimated directly from single crystal elastic constants using the Voigt-Reuss-Hill (VRH) approximation, according to work [31]

$$B = \frac{1}{2}(B_V + B_R), G = \frac{1}{2}(G_V + G_R) \quad (19)$$

$$E = \frac{9GB}{3B+G}, v = \frac{3B-E}{6B} = \frac{E-2G}{2G} \quad (20)$$

Where *V* indicates the model of Voigt and *R* denotes the Reuss model. The Voigt model provide the upper assessments of values of mechanical modulus, while the Reuss model provide their lower assessments. The moduli of hexagonal crystals are calculated according to work [32]:

$$B_V = \frac{1}{9}(2(C_{11} + C_{12}) + 4C_{13} + C_{33}) \quad (21)$$

$$B_R = \frac{(C_{11}+C_{12})C_{33}-2C_{13}^2}{C_{11}+C_{12}+2C_{33}-4C_{13}} \quad (22)$$

$$G_V = \frac{1}{30}(C_{11} + C_{12} + 2C_{33} - 4C_{13} + 12C_{44} + 12C_{66}) \quad (23)$$

$$G_R = 15/(14S_{11} + 4S_{33} - 8S_{13} - 10S_{12} + 6S_{44}) \qquad (24)$$

Where $C_{ij}$ are elastic constants and $S_{ij}$ are components of the compliances tensor, which is the inverse matrix of $C_{ij}$ strain tensor. The obtained elastic compliances $S_{ij}$ of single crystal of Laves phase $Fe_2Mo$ are listed in Table 5.

**Table 5**
Ground state elastic compliances $S_{ij}$ (TPa$^{-1}$) of single crystal of Laves phase $Fe_2Mo$.

| $S_{11}$ | $S_{12}$ | $S_{13}$ | $S_{33}$ | $S_{44}$ |
|---|---|---|---|---|
| 2.724 | -0.737 | -1.191 | 7.107 | 8.793 |

Calculated by VRH approximation method modules of polycrystalline $Fe_2Mo$ aggregate are shown in Table 6. The Young's modulus $E$ quantifies the relationship between tensile stress σ and axial strain ε. The bulk modulus $B$ is a measure of resistance to volume change by applied pressure. Shear modulus $G$ is a measure of resistance to reversible deformations upon shear stress. Poisson's ratio defines the ratio of transverse strain to the axial strain [33].

**Table 6**
Ground state elastic modulus (GPa) and Poisson's ratio $v$ of polycrystalline aggregate of $Fe_2Mo$

| $B_V$ | $B_R$ | $B$ | $G_V$ | $G_R$ | $G$ | $E$ | $v$ | $B/G$ |
|---|---|---|---|---|---|---|---|---|
| 228.9 | 224.1 | 226.5 | 135.5 | 110.1 | 122.8 | 312.0 | 0.27 | 1.84 |

A concept of material plasticity includes, among others, brittleness and ductility. In order to have a measure to separate between brittle and ductile behaviour of materials, in the work [33] was proposed the ratio of bulk modulus to shear modulus ($B/G$). A high value of $B/G$ ratio means the ductile behaviour of materials, whereas a low value associated with the brittle behaviour. The critical value which separate ductile and brittle material was suggested as 1.75. It was suggested in [34] that Poisson's ratio can also represent such a separation with a critical value $v = 0.26$, so a material with $v > 0.26$ is referred to as having a ductile nature.

Thus, calculated values of $B/G$ ratio and Poison's ratio predict that Laves phase $Fe_2Mo$ is a material with ductile behaviour at T = 0 K.

*3.1.3 Thermal properties*

The thermodynamic and physical properties of compounds like elastic constants, vibration entropy, thermal expansion, specific heat, and melting temperature can be calculated through the Debye temperature $\theta_D$.

In this work the Debye temperature $\theta_D$ was calculated through polycrystalline bulk $B$ and shear $G$ modulus by method described in the work [35]. To do this, in first the $v_s$ shear and $v_l$ longitudinal elastic wave velocities were calculated using the polycrystalline bulk modulus $B$ and shear modulus $G$ as follows:

$$V_s = \left(\frac{G}{\rho}\right)^{1/2}, \quad V_l = \left(\frac{(B+4G/3)}{\rho}\right)^{1/2} \qquad (25)$$

Where $B$ and $G$ are the bulk and shear modulus, $\rho$ is a density of the compound. Then, the averaged sound velocity $V_m$ in polycrystalline material was calculated as follows:

$$V_m = \left[\frac{1}{3}\left(\frac{2}{V_s^3} + \frac{1}{V_l^3}\right)\right]^{-1/3} \qquad (26)$$

And, the Debye temperature $\theta_D$ was calculated by the equation as follows [35]

$$\theta_D = \frac{h}{k_B}\left[\frac{3n}{4\pi}\left(\frac{N_A\rho}{M}\right)\right]^{1/3} V_m \qquad (27)$$

Where $h$ and $k_B$ are the Plank and the Boltzmann constants, $N_A$ is Avogadro constant, $\rho$ is the density, $n$ is the number of atoms per formula unit, $M$ is the molecular weight of the solid.

The calculated results of elastic wave velocities ($v_s$, $v_l$, $V_M$) and $\theta_D$ for Fe$_2$Mo are listed in Table 8. The obtained density of Fe$_2$Mo compound is 9.38 (g/sm$^3$). According to the empirical rule, the higher the Debye temperature is, the better the thermal conductivity of the material. So, the predicted $\theta_D$ can be a measure of Fe$_2$Mo thermal conductivity. Another important feature of $\theta_D$ is the using in calculations of Gibbs potential depending on temperature in the frame of quasi-harmonic Debye - Grüneisen approach. The calculated value of Debye temperature $\theta_D$ at T=0K was used in this work for the calculation of $C_{ij}(T)$ at finite-temperatures which are listed in Table 7. As far as is known, there are no experimental or theoretical value of Debye temperature and sound velocities available for the Fe$_2$Mo phase. Thus, these results are of academic interest and may be usable for future works.

The velocities of sound waves propagating in a hexagonal crystal in different directions were calculated in accordance with [36, 37]. The velocities of elastic waves propagating along [001] direction were calculated as follows:

$$V_L = \left(\frac{C_{33}}{\rho}\right)^{1/2}, \quad V_{S1} = V_{S2} = \left(\frac{C_{44}}{\rho}\right)^{1/2} \qquad (28)$$

And the velocities of elastic waves propagating along [100] direction were given as:

$$V_L = \left(\frac{C_{11}}{\rho}\right)^{1/2}, \quad V_{S1} = \left(\frac{C_{66}}{\rho}\right)^{1/2}, \quad V_{S2} = \left(\frac{C_{44}}{\rho}\right)^{1/2} \qquad (29)$$

The calculated results of anisotropic elastic waves velocities are shown in Table 8. The fastest propagation of elastic waves was found along the [100] direction by longitudinal waves. Prediction of elastic wave velocities in different directions of the crystal may be useful for experimental studies.

*3.1.4 Calculation method layout*

In order to investigate the stability of Fe$_2$Mo with different lattice parameters and predict a path of thermal expansion, a new approach was applied. It consists of the following. On a coordinate plane (*a*, *c*) one puts a point with coordinate ($a_0$,$c_0$), it is the relaxed parameters of Fe$_2$Mo lattice obtained in this work by DFT at T = 0K and listed in Table 2. Through this point one draws a family of paths (or line segments). Here, as an example, ten paths are shown in Figure 3. If one will move an imaginary point with coordinates (*a*, *c*) along the *n0* path it could be seen that *c/a* = 1.664 ratio remains constant, this is the path of the isotropic expansion and contraction. The *n1* path is passing through coordinates of the experimental point obtained at T = 1073 K after 1500 hours of annealing [26] and shown in Figure 3 by a red triangle. Along the *n9* path the parameter *a-* remains constant. By increasing the family of paths and the number of points on them, and calculating the energy at each point, one can obtain the energy surface $E_{tot}(a, c)$. Along each *ni* paths the lattice parameters *a-* and *c-* have a functional dependence of $c = f^{ni}(a)$ with each other. It's necessary to note, that each point laying on each *ni* path corresponds the unequivocally defined volume $V(a,c) = (\sqrt{3}/2) \cdot a^2 c$ of Fe$_2$Mo lattice. Thus, moving along the *ni* paths, the total energies $E^{ni}_{tot}(V)$ can be calculated. These calculations are shown in Figure 4. To simplify the calculation of thermodynamic properties of compounds, instead of dealing with the energy surface $E_{tot}(a, c)$, one can calculate the energies $E^{ni}_{tot}(V)$ along such *ni* paths. This technique allows us to reduce the problem to a one-dimensional case and consider the energy for each individual path taken as a function of one variable, the volume *V*. That makes it possible to avoid the need for differentiation with respect to several variables. Therefore, it allows to use the Birch-Murnaghan equation [11] and apply the quasi-harmonic Debye- Grüneisen model [8] for each path to simulate the effect of temperature $T$ on the thermodynamic properties of compounds by calculating free energies $F^{ni}(V,T)$ along each path.

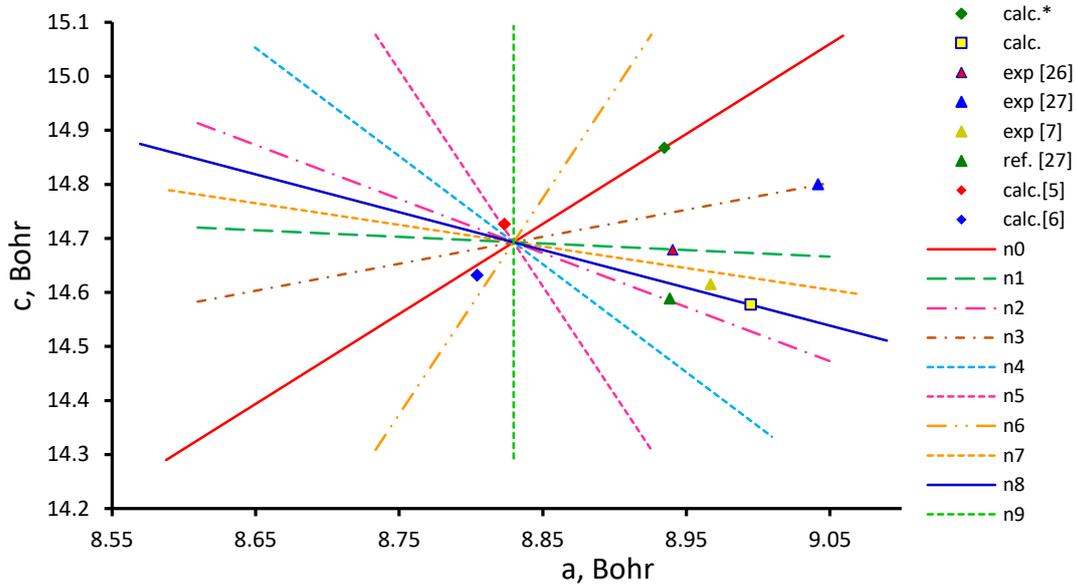

**Figure 3**. The coordinate plane (a, c) and ten paths (line segments) $n0 \div n9$ where the intersection point of all paths corresponds to the lattice parameters ($a_0$, $c_0$) of the Fe$_2$Mo calculated in this work by DFT at T = 0 K. The experimental point obtained at T=1073K after 1500 hours of annealing [26] is shown by the red triangle. The blue triangle is experimental point with composition of (in at. %) 65Fe25Mo10Ta obtained at T= 1073K [27]. The other available experimental points obtained at room temperatures are olive and green triangles taken from works [7, 27]. The results calculated in this work using the quasi-harmonic Debye - Grüneisen theory at T = 1073K with and without accounting magnetic entropy are shown by a yellow square and a green diamond, respectively. The other theoretical calculations of Fe$_2$Mo parameters calculated by DFT at T = 0 K [5, 6] are shown with red and blue diamonds.

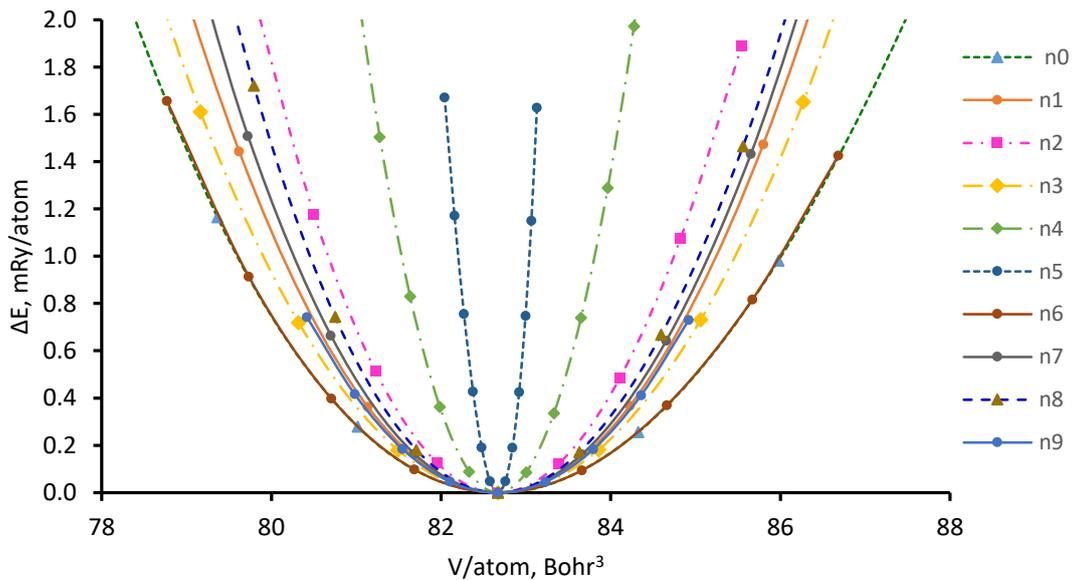

**Figure 4**. Total energies $E^{ni}_{tot}(V)$ of Laves phases Fe$_2$Mo calculated along $n0 \div n9$ paths.

Comparing the temperature dependences of free energies $F^{ni}(V,T)$ calculated by formula (1) for the *ni* paths, and, if necessary, adding more paths for such comparison procedure, one can obtain the most energetically favorable $F^{ni}(V,T)$ function with lowest energy. Each such *ni* path can be considered as a pseudo-phase and they can be compared between themselves like it does in the CALPHAD method. Having found, as a result of this procedure, the energy $F^{ni}(V,T)$ of the most stable pseudo-phase, one can determine the *ni* path as the trajectory of the thermal expansion and contraction path at heating - cooling and calculate thermodynamic properties of a compound along this path. It necessary to note that, by adding more paths and comparing the free energies calculated along these paths, one can improve the accuracy of the calculations. As a result, the thermal expansion path may turn out to be non-linear.

### *3.2. Compound stability at finite temperatures*

### *3.2.1. Debye temperature calculations*

In order to investigate an influence of vibrational entropy on the $Fe_2Mo$ stability at different temperatures the analyses of Debye temperatures as functions of volume calculated for different paths *n0 ÷ n9* was provided.

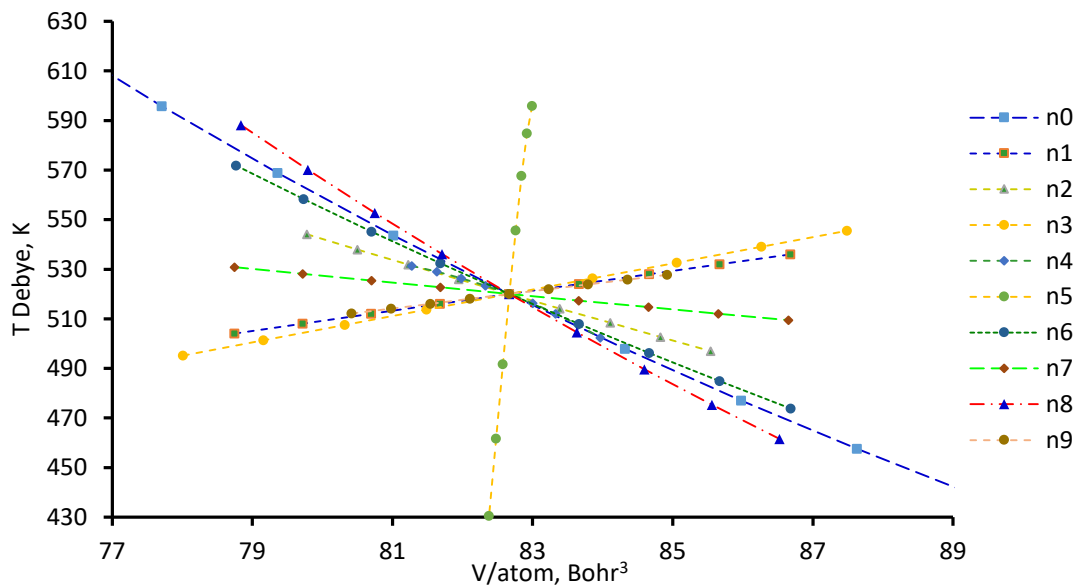

**Figure 5**. Debye temperatures $\theta_D(V)$ of Laves phases $Fe_2Mo$ calculated along *n0 ÷ n9* paths.

Debye temperatures $\theta_D(V)$ calculated for different paths are shown in Figure 5. From those graphs one could conclude that according the Debye – Grüneisen model the *n0, n2, n4, n6, n7, n8* paths should be considered as perspectives because theirs curves have negative overall slopes and will contribute more to the vibrational energy than other paths. Due to the peculiarity of the nature of the hexagonal crystal lattice the compounds, belonging to the *n5* path, have small differences in their volumes, which leads to a strong redistribution of magnetic moments (as shown in Figures 6-9) with small changes in volume and, as a consequence, to a non-physical result. This feature is reflected in the atypical asymmetry of the total energy, as shown in Figure 4, where its curvature is higher with expansion and lower with contraction of volume that leads to the negative thermal expansion effect, and as a follow in a large positive slope of the $\theta_D(V)$ curve shown in Figure 5. Therefore, the *n5* path was not considered in this analysis and is presented here for reference only.

## 3.2.2. Distribution of local magnetic moments over sublattices in the Fe₂Mo

As was shown in the works [9, 10] the accounting local magnetic moments distributed on sublattices gets an essential contribution about 30% to the free energy. So, in order to investigate influence of magnetic entropy on the Fe₂Mo stability with different lattice parameters the dependencies of local magnetic moments, in Bohr magnetons ($\mu_B$), on the volumes were obtained along $n0 \div n9$ paths and shown in Figures 6-9. Studying the magnetic properties of Fe₂Mo compound in more details it was found that at equilibrium volume $V_0$, that associates with T = 0 K, the largest contribution to the magnetic entropy is made by Fe atoms located on the *6h* sub-lattice, about 63%. The Fe atoms located on the *2a* sub-lattice contribute about 21%, the Mo (*4f*) atoms about 2% and the interstitial space contributes 14%. While at the 83.4÷88.0 (a.u.³/atom) range of volume (that associates with T = 500 ÷ 1000 K) the most contribution to the magnetic entropy comes from Fe (*6h*) atoms and interstitial space, about 55% and 35% accordingly, and the rest comes from Fe (*2a*) 4% and Mo (*4f*) atoms 6%. A significant change in the magnetic ordering along the *n8* path was found with the expansion of the volume starting from $V$ = 83.8 (a.u.³/atom), which corresponds to the temperature T = 547 K.

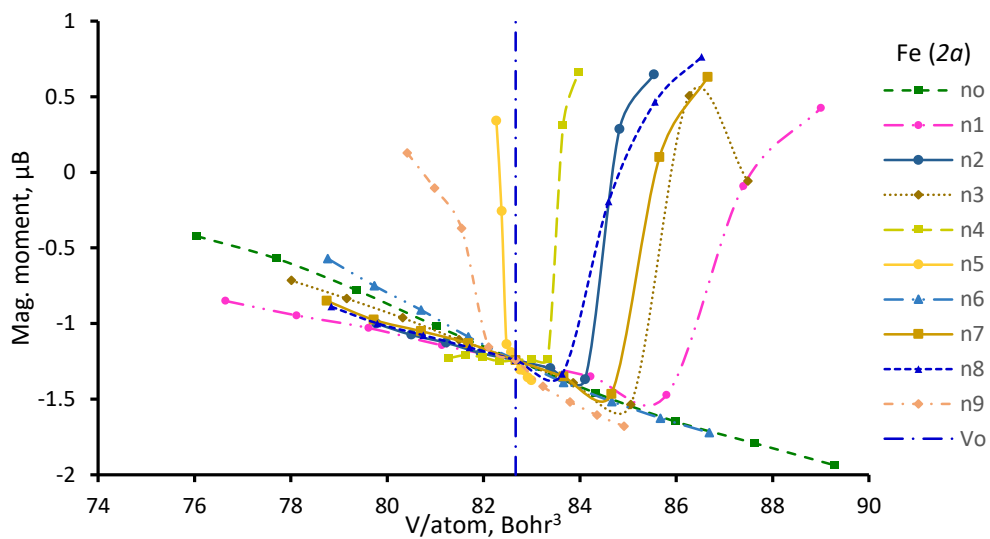

**Figure 6**. Distribution of local magnetic moments ($\mu_B$) of Fe atoms on the first sub-lattice (*2a*) of the Laves phases Fe₂Mo obtained along $n0 \div n9$ paths; $V_0$ – equilibrium volume.

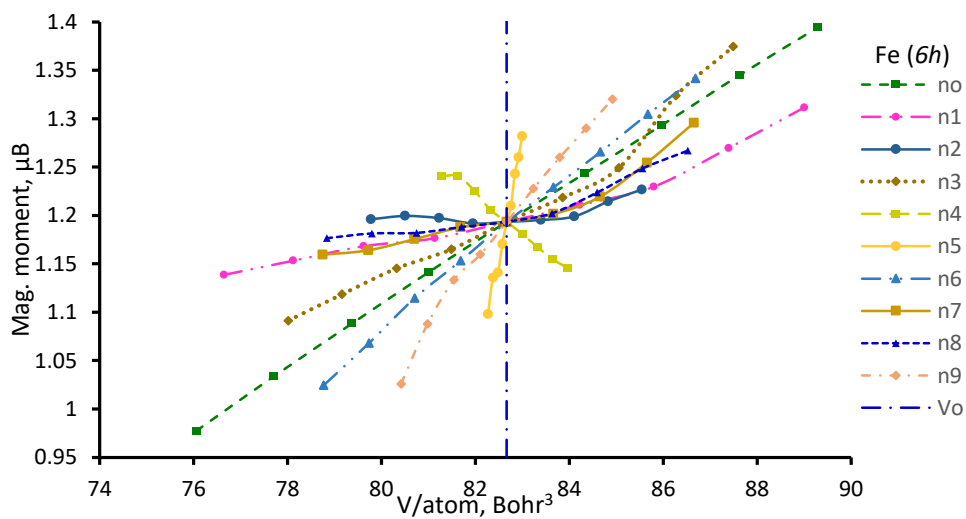

**Figure 7**. Distribution of local magnetic moments ($\mu_B$) of Fe atoms on the second sub-lattice (*6h*) of the Laves phases Fe₂Mo obtained along $n0 \div n9$ paths; $V_0$ – equilibrium volume.

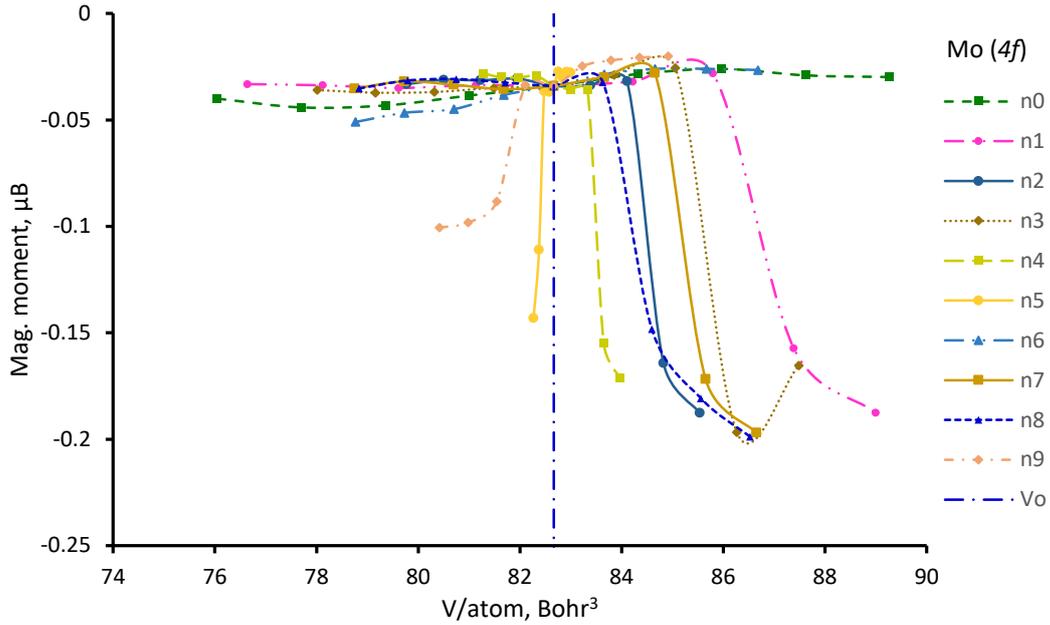

**Figure 8**. Distribution of local magnetic moments ($\mu_B$) of Mo atoms on the third sub-lattice (*4f*) of the Laves phases Fe$_2$Mo obtained along *n0* ÷ *n9* paths; V$_0$ – equilibrium volume.

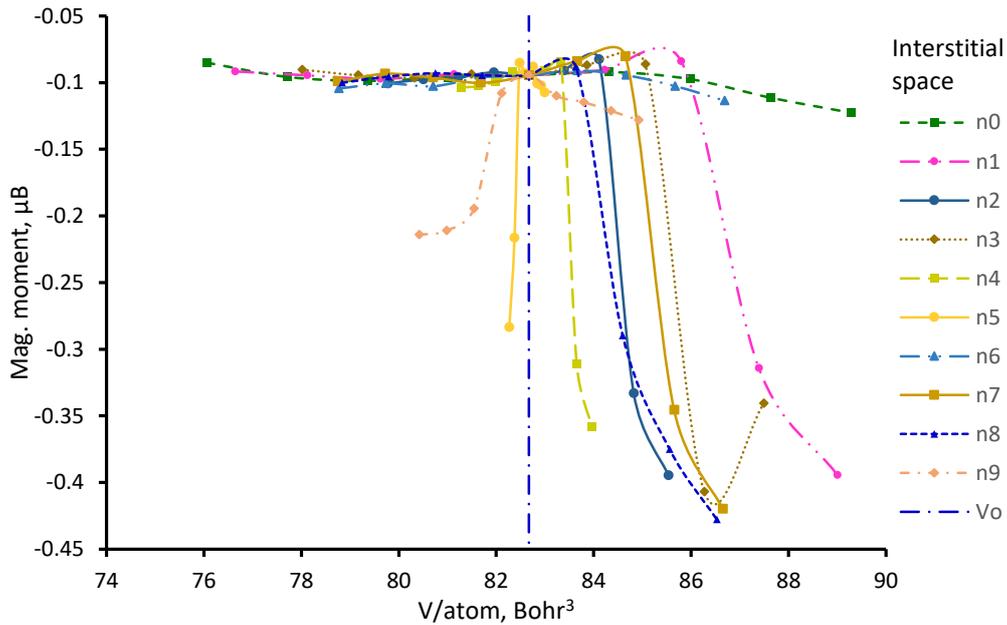

**Figure 9**. Distribution of local magnetic moments ($\mu_B$) in the interstitial space of the Fe$_2$Mo lattice obtained along *n0* ÷ *n9* paths; V$_0$ – equilibrium volume.

Along other *n4*, *n2*, *n7*, *n3*, and *n1* paths, noticeable changes in the magnetic entropy begin with an increase in volumes starting from 83.4, 84.3, 84.8, 85.2 and 86.0 (a.u.$^3$/atom) respectively. Along *n0* and *n6* paths, no significant change in the magnetic ordering is observed with increasing volume. But, along the *n9* path, an increase in the entropy occurs with a decrease in volume, probably, this may be due to the peculiarity of the structure of the hexagonal crystal lattice. And just as with the data obtained from the *n5* path, the results of *n9* should be treated carefully.

*3.2.3. Calculations of free energies*

To implement the proposed new method of searching for the optimal path of thermal expansion of $Fe_2Mo$, the Helmholtz free energies $F^{ni}(V,T)$ for *ni* paths are calculated by (1) at finite temperatures. An example of such a calculation of free energies $F^{n8}(V,T)$ for the case of *n8* path, for temperatures from T = 100K to T = 1100K, is shown in Figure 10. This Figure also shows the total energy $E^{n8}_{tot}(V)$ calculated by DFT at T = 0K along *n8* path, which is shown by the red dashed line.

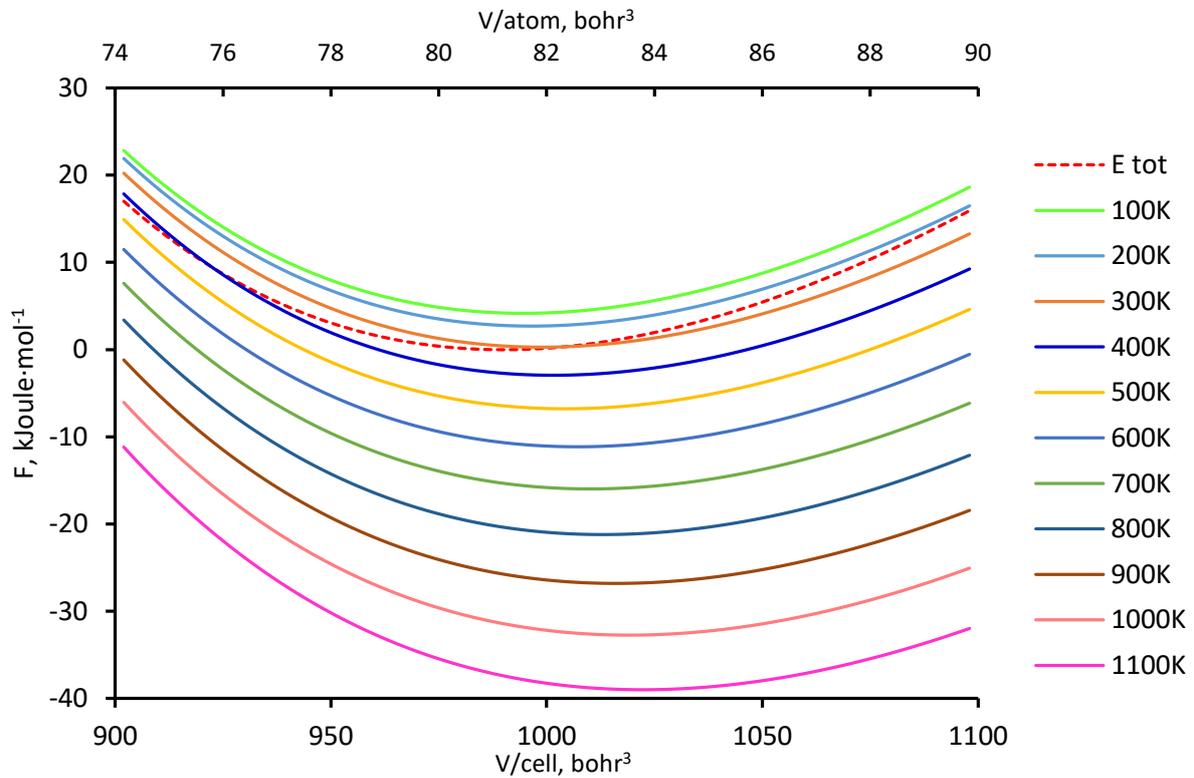

**Figure** 10. The Helmholtz free energy curves $F^{n8}(V,T)$ of Laves phase $Fe_2Mo$ calculated at a number of temperatures along the *n8* path. The red dotted line is the $E^{n8}_{tot}(V)$.

For each curve of $F^{n8}(V,T)$ calculated at particular temperature there is a minimum corresponding to an equilibrium volume $V_0$. Figure 10 shows that with increasing temperature, the equilibrium volume of the compound increases its value.

The free energies $F^{ni}(T)$ for each *ni* path are calculated by (1) at the equilibrium volumes $V_0$, which correspond to the minimum value of $F^{ni}(V,T)$ function and the current temperature T. In the same way, all energy contributions to the free energy are taken into account in the calculations: electronic, vibrational and magnetic, which are described by (2-15).

The curves of the electronic contributions $F^{ni}_{el}(T)$ to the Helmholtz free energy calculated by (2 – 4) along *n0* ÷ *n9* paths are shown in Figure 11. From this Figure follows that the electronic contribution to the free energy is greater from the *n8* path than from the others.

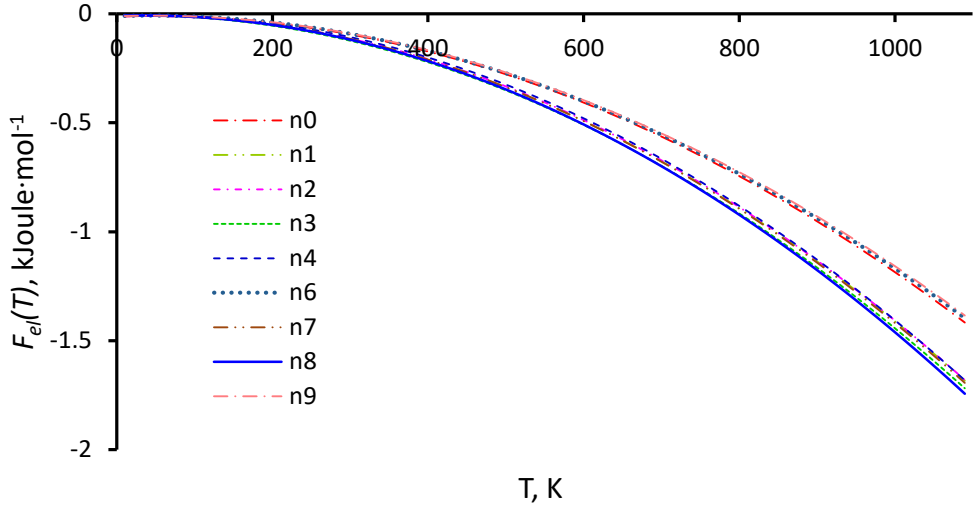

**Figure** 11. Electronic energies $F^{ni}_{el}(T)$ of Laves phase $Fe_2Mo$ calculated along $n0 \div n9$ paths.

The distributions of magnetic moments on sublattices in the $Fe_2Mo$ compound shown in Figures 6 - 9 were taken into account by (14). The magnetic entropy $S^{ni}_{mag}(V_0)$ for each $ni$ path was calculated by (14) at equilibrium volume $V_0$ which corresponds to the minimum value of $F^{ni}(V,T)$ and particular temperature $T$. So, the magnetic entropies $S^{ni}_{mag}(V_0)$ calculated at equilibrium volumes and multiplied by $-T$ are shown in Figure 12 for different $ni$ paths. From Figure 12 one could conclude that the contribution to the magnetic entropy is more significant from $n8$ path then from others.

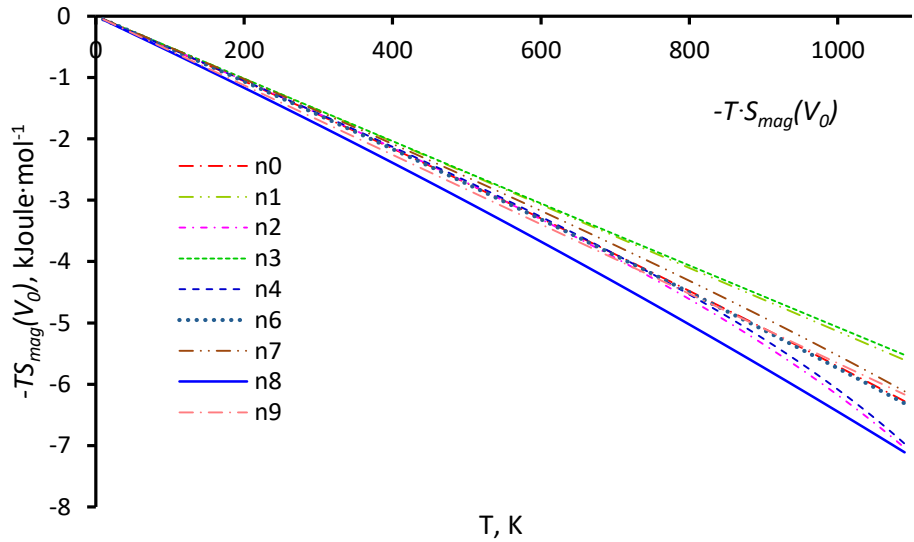

**Figure** 12. Magnetic entropies $-T \cdot S^{ni}(V_0)$ of Laves phase $Fe_2Mo$ calculated along $n0 \div n9$ paths.

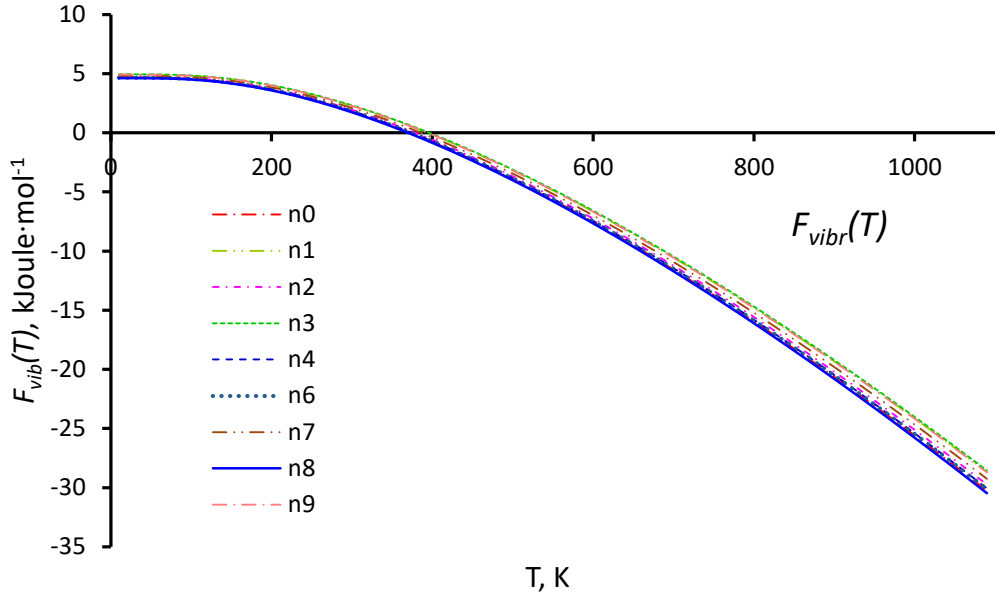

**Figure** 13. Vibrational energies $F_{vib}(T)$ of Laves phase Fe$_2$Mo calculated for $n0 \div n9$ paths.

The vibrational energy contributions $F^{ni}_{vib}(T)$ to the free energy were calculated by (5 – 10) for different $n0 \div n9$ paths and shown in Figure 13. From Figure 13 it follows that the $n8$ path is energetically favorable one again.

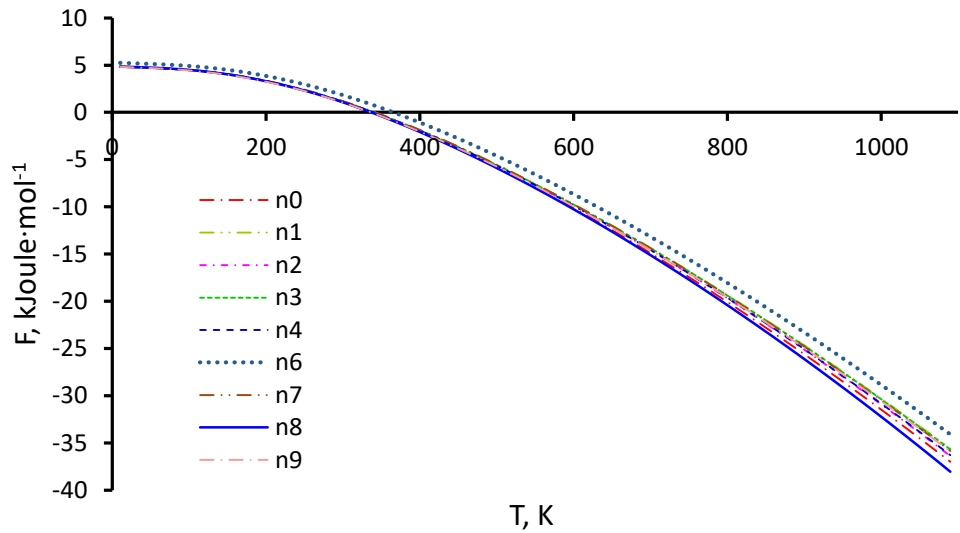

**Figure** 14. The Helmholtz free energy curves $F^{ni}(T)$ of Laves phase Fe$_2$Mo calculated along $n0 \div n9$ paths.

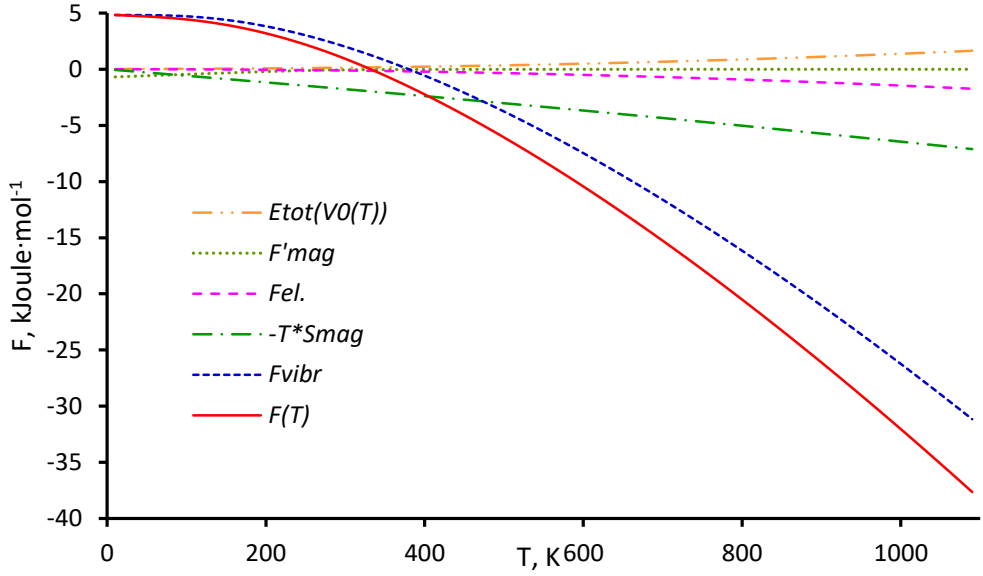

**Figure** 15. The Helmholtz free energy $F^{n8}(T)$ of Laves phase $Fe_2Mo$ together with its energy contributions (i.e. total, magnetic, electronic, magnetic entropy and vibrational) calculated as the temperature dependence along the *n8* path.

The free energies $F^{ni}(T)$ consisting of these energy contributions and calculated by (1) for different paths are shown in Figure 14. It's necessary to point out that the configurational entropy $S_{conf}(T)$ calculated by formula (15) in the case of a compound with stoichiometry, such as $Fe_2Mo$, gives the same energy contribution for all paths *ni*, and does not affect comparative analysis. Therefore, analyzing these curves, one can conclude that *n8* path (or this pseudo-phase) is the most stable or energetically favorable among others, it means that during the heating up to the T=1100K the lattice parameters of the $Fe_2Mo$ compound will increase from ($a_0$, $c_0$) towards values according the law associated with the *n8* path or the line segment shown in Figure 3. In other words, this is the path of thermal expansion and contraction of $Fe_2Mo$.

The free energy $F^{n8}(T)$ of Laves phase $Fe_2Mo$ calculated along *n8* path and its constituents, namely, $E^{n8}_{tot}(V_0(T))$ total energy, $F'^{n8}_{mag}(T)$ magnetic energy (12), $F^{n8}_{el}(T)$ electronic (2 – 4), $F^{n8}_{vibr}(T)$ vibrational (5 – 10) and $-T \cdot S^{n8}_{mag}(T)$ magnetic entropy (14), are shown in Figure 15.

In addition, analysis of energy contributions calculated along $n0 \div n9$ paths shows that the other energetically closest path to the stable *n8* is the *n0* path, as shown in Figure 14, and if one ignores the magnetic entropy the stable path would be the *n0* path. Therefore, this calculation shows that disregarding the distribution of magnetic moments over the $Fe_2Mo$ sublattices, the energetically more stable path was *n0* and predicted parameters at T=1073K would be $a = 4.728$ Å and $c = 7.868$ Å, whereas allowing the magnetic entropy, the path shifts to the *n8* trajectory and the parameters are $a = 4.760$ Å and $c = 7.714$ Å, these results are listed in Table 2 and Figure 3 for reference. Therefore, the influence of magnetic entropy on the stability of $Fe_2Mo$ cannot be ignored.

### 3.2.4. Calculations of thermodynamic and physical properties

Knowing the temperature dependence of free energy $F^{n8}(T)$ of the most energetically stable *n8* path among others, one can calculate thermodynamic and physical properties of $Fe_2Mo$. The calculated curve of volume expansion $V(T)$ as a temperature function is shown in Figure 16, and for the sake of comparison the experimental data obtained at T = 1073K after 1500 hours of annealing [26] is shown here as well, but it belongs to the *n1* path, as shown in Figure 3, which is not energetically favorable according to this calculation and it is away from the *n8* thermal expansion–contraction path.

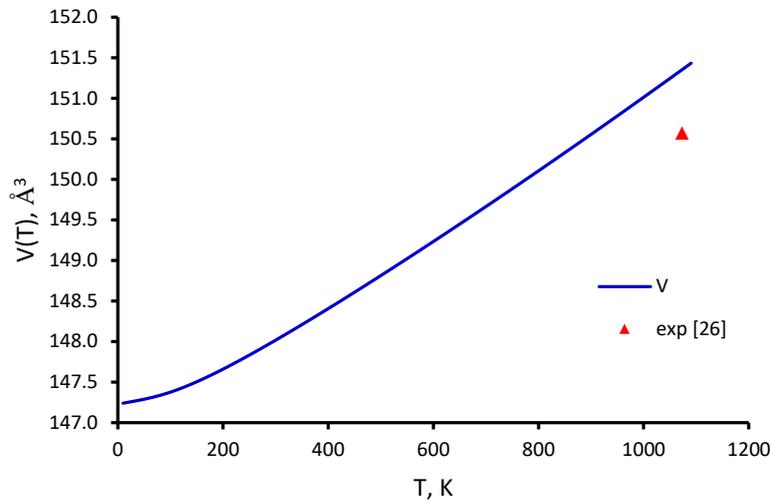

**Figure** 16. Calculated volume, a solid line, *V(T)* (in Å$^3$) of Laves phase Fe$_2$Mo vs temperature in comparison with the experimental data obtained at T = 1073 K after 1500 hours of annealing [26] shown by the red triangular point.

The Curie temperature, $T_c$, was evaluated by (13). The difference between the ground state total energies of PM and FM Fe$_2$Mo has been calculated at the equilibrium atomic volumes of the FM and PM states. The results of the assessment showed that the value of the $T_c$ is 299K.

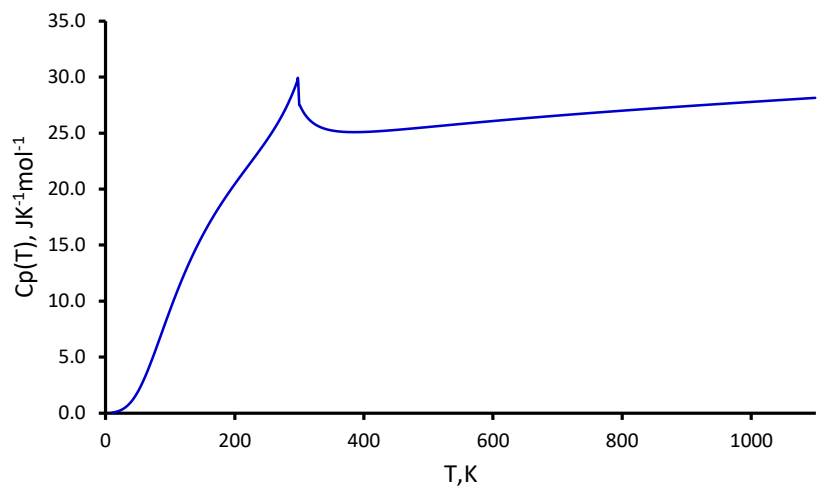

**Figure** 17. The calculated heat capacity $C_p(T)$ of Laves phase Fe$_2$Mo calculated along the thermal expansion – contraction path.

The temperature dependence heat capacity *Cp(T)* calculated along the *n8* thermal expansion path is shown in Figure 17, the jump at T = 299 K corresponds to the ferro-/ paramagnetic phase transition of the Fe$_2$Mo according to the applied model [16, 17]. As far as is known, there aren't any available experimental heat capacity and magnetometry studies on this compound, so the exact value of the Curie temperature for the Fe$_2$Mo is a focus for further studies.

The elastic constants $C_{ij}(T)$ together with the modulus $B(T)$, $G(T)$ and $E(T)$, which we obtained by VRH approximation (19 – 24) and calculated along the *n8* thermal expansion path by the distortion matrices $D_i$, are shown in Figure 18. The elastic constants $C_{ij}$, modulus $B$, $G$ and $E$, Poison's $v$ and $B/G$ ratios calculated at number of temperatures are listed in Table 7. As one could see in Figure 18 and Table 7, that the elastic constants of $Fe_2Mo$ satisfy the mechanical stability criteria (18) and as follow this compound is remained mechanically stable during heating.

The calculated values of $B/G$ ratio and Poison's ratio, according the work [33, 34], predict that Laves phase $Fe_2Mo$ is a material with ductile behaviour at low temperatures up to about T=500 ÷ 600K and at higher temperatures it begins to show its brittle nature.

The Debye temperatures, average, shear and longitudinal elastic wave velocities calculated by (25-29) at finite temperatures along the *n8* thermal expansion path are shown in Table 8. From these data, it follows that the Debye temperature at heating up to 1000 K changes its value from 520 K toward 480 K, and the $Fe_2Mo$ compound, according to the rule of thumb, decrease its thermal conductivity.

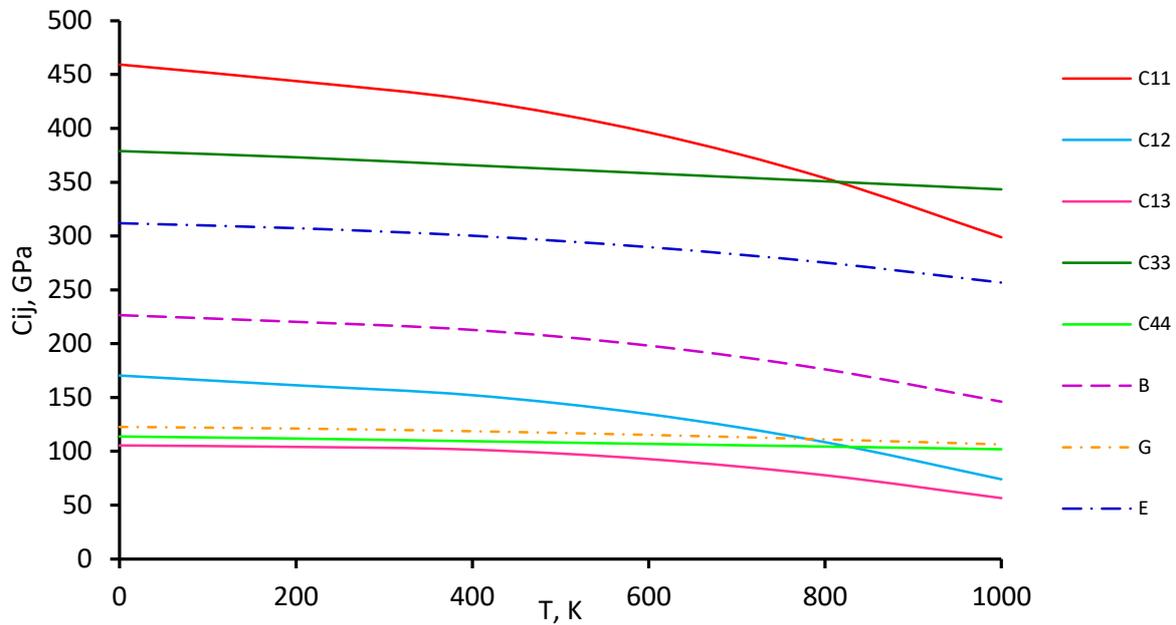

**Figure** 18. Elastic constants $C_{ij}$ and modules $B$, $G$, $E$ of Laves phase $Fe_2Mo$ calculated as functions of temperature along the thermal expansion – contraction path.

**Table 7**
The elastic constants $C_{ij}$, modulus (GPa) and Poisson's ratio $v$ calculated by distortion matrices at number of temperatures along the thermal expansion – contraction path of $Fe_2Mo$.

| T, K | $C_{11}$ | $C_{12}$ | $C_{13}$ | $C_{33}$ | $C_{44}$ | B | G | E | v | B/G |
|---|---|---|---|---|---|---|---|---|---|---|
| 0 | 459.3 | 170.4 | 105.5 | 379.0 | 113.7 | 226.5 | 122.8 | 312.0 | 0.27 | 1.84 |
| 200 | 443.9 | 161.3 | 104.1 | 373.2 | 111.8 | 220.3 | 121.2 | 307.3 | 0.27 | 1.82 |
| 400 | 426.3 | 152.1 | 101.6 | 365.7 | 109.3 | 212.8 | 118.7 | 300.3 | 0.26 | 1.79 |
| 600 | 396.2 | 134.5 | 92.8 | 358.2 | 106.8 | 198.2 | 115.3 | 289.7 | 0.26 | 1.72 |
| 800 | 353.7 | 108.4 | 77.8 | 350.7 | 104.3 | 176.1 | 111.1 | 275.3 | 0.24 | 1.59 |
| 1000 | 298.8 | 73.9 | 56.6 | 343.3 | 101.8 | 146.0 | 106.3 | 256.7 | 0.21 | 1.37 |

**Table 8**
The calculated average ($V_m$), shear ($v_s$) and longitudinal ($v_l$) elastic wave velocities (m/s); predicted Debye temperatures $\theta_D$ (K), and elastic wave velocities along [001] and [100] directions (m/s) at number of temperatures along the thermal expansion – contraction path of Laves phase Fe$_2$Mo.

| T,K | $v_s$ | $v_l$ | $V_m$ | $\theta_D$ | [001] | | [100] | | |
|---|---|---|---|---|---|---|---|---|---|
| | | | | | $V_l$ | $V_s$ | $V_l$ | $V_{s1}$ | $V_{s2}$ |
| 0 | 3618 | 6449 | 4026 | 520 | 6356 | 3481 | 6997 | 3924 | 3481 |
| 200 | 3595 | 6380 | 3999 | 516 | 6307 | 3452 | 6879 | 3881 | 3452 |
| 400 | 3557 | 6289 | 3956 | 511 | 6243 | 3413 | 6741 | 3823 | 3413 |
| 600 | 3505 | 6124 | 3895 | 503 | 6179 | 3374 | 6499 | 3735 | 3374 |
| 800 | 3441 | 5878 | 3815 | 493 | 6114 | 3335 | 6140 | 3616 | 3335 |
| 1000 | 3366 | 5538 | 3719 | 480 | 6049 | 3295 | 5643 | 3462 | 3295 |

*3.2.5. Discussion*

In summary, by applying the new method of searching a path of thermal expansion–contraction and taking into account the magnetic entropy in calculating the free energy of Fe$_2$Mo makes it possible to obtain the most energetically favorable direction of thermal expansion, it's the *n8* path. Without accounting the distribution of magnetic moments of atoms, the obtained direction will correspond to the isotropic expansion–contraction path, this result is shown by green diamond (and noted as calc.*) in Figure 3 and corresponds to the *n0* path. But all experimental data show that this is not the case, the experimental points are located far from the direction of the isotropic expansion–contraction *n0* path, as shown in Figure 3. For example, the experimental point obtained at T=1073K after 1500 hours of annealing in the work [26] lies on the *n1* path, shown in Figure 3 by red triangle. Another data of the Fe$_2$Mo lattice parameters taken from [27] where they are marked as taken from a reference and obtained at room temperature, this point lies on the *n2* path and marked by the green triangle. By alloying of an additional element of 10 at. % Tantalum into the Fe$_2$Mo compound leads to the displacement of the experimental point obtained at T = 1073 K [27] to the *n3* path it shown by the blue triangle. The lattice parameters of Fe$_2$Mo taken from [7] which are experimentally obtained at room temperature lay almost exactly on the *n8* path as shown in Figure 3 by the olive triangle. Such a scatter of experimental data may be explained by slow diffusion and the presence of other phases in the Fe - Mo system, as discussed above. Therefore, the experimentally obtained parameters of the Fe$_2$Mo lattice may be considered as showing that the newly proposed method can correctly predict the path of thermal expansion. It should be noted that the thermal path may not be a line but a curve. Here the calculations were limited along the straight line paths. The search of more exact curved path is a subject for further research. Thus, the role of magnetic subsystem in the free energy of Fe$_2$Mo is essential.

**4. Conclusions**

The first-principles calculations have been carried out to calculate electronic, vibrational and magnetic free energy contributions for the Laves phases Fe$_2$Mo. The new method of finding a thermal expansion and contraction path of compounds is developed. This approach makes it possible to reduce the problem to a one-dimensional case and avoid the need for differentiation in many variables for calculating thermodynamic functions. The optimal thermal expansion path of Fe$_2$Mo is obtained by comparison the free energies calculated along different paths. The elastic constants $C_{ij}(T)$, bulk modulus, the lattice volume expansion $V(T)$, heat capacity $C_p(T)$, elastic sound velocities and Debye temperature $\theta_D(T)$ of Fe$_2$Mo are predicted along this path. The influence of the magnetic entropy on stability the intermetallic compound is shown. If treats the Fe$_2$Mo without magnetic entropy then a stable thermal path will be of an isotropic expansion – contraction one, but this theoretical calculation and the experimental results show that the path of thermal expansion lays at another direction. This proves that the newly developed method can predict the thermal expansion path. Electronic, vibrational and magnetic entropy contributions must be taken into account in a similar way when calculating thermodynamic properties of intermetallic compounds containing materials with magnetic properties.

**Acknowledgments**

The research was financially supported by the Russian Foundation for Basic Research as a part of scientific project № 19-03-00530.